\documentclass[11pt,letterpaper]{article}
\usepackage[utf8]{inputenc}
\usepackage[T1]{fontenc}
\usepackage[margin=0.75in]{geometry}
\usepackage{mathptmx}
\usepackage{amsmath,amssymb,graphicx,needspace,titlesec}
\usepackage{xcolor}
\usepackage[colorlinks=true,urlcolor=blue,linkcolor=blue,citecolor=blue]{hyperref}
\hypersetup{pdftitle={The effect of limb kinematics on the speed of a legged robot on granular media},pdfauthor={Chen Li; Paul B. Umbanhowar; Haldun Komsuoglu; Daniel I. Goldman}}
\titleformat{\section}{\fontsize{13}{16}\selectfont\bfseries}{\thesection}{.65em}{}
\titlespacing*{\section}{0pt}{14pt}{7pt}
\renewenvironment{abstract}{\section*{Abstract}}{\par}
\let\originalthebibliography\thebibliography
\let\endoriginalthebibliography\endthebibliography
\renewenvironment{thebibliography}[1]{\originalthebibliography{#1}\fontsize{9.3}{11.625}\selectfont\setlength{\itemsep}{3pt}\setlength{\parskip}{0pt}}{\endoriginalthebibliography}
\makeatletter
\renewcommand{\@biblabel}[1]{#1.}
\makeatother
\newsavebox{\manuscriptimage}
\newcommand{\manuscriptfigure}[4]{%
  \par\begingroup
  \sbox{\manuscriptimage}{\includegraphics[#2]{#3}}%
  \Needspace{\dimexpr\ht\manuscriptimage+\dp\manuscriptimage+42pt\relax}%
  \noindent\makebox[\linewidth][c]{\hypertarget{fig#1}{}\usebox{\manuscriptimage}}\par
  \vspace{2pt}%
  \refstepcounter{figure}\label{figure#1}%
  {\fontsize{9.5}{12.35}\selectfont\noindent\textbf{Figure \thefigure.} #4\par}%
  \endgroup\vspace{8pt}%
}

\titleformat{\subsection}{\fontsize{11.5}{14}\selectfont\bfseries}{\thesubsection}{.65em}{}
\titlespacing*{\subsection}{0pt}{10pt}{5pt}
\begin{document}
\fontsize{11.2}{15.23}\selectfont
{\centering\fontsize{18}{21.6}\selectfont\bfseries The effect of limb kinematics on the speed of a legged robot on granular media\par}
\vspace{8pt}
{\centering\fontsize{11}{15.4}\selectfont Chen Li$^{1}$, Paul B. Umbanhowar$^{2}$, Haldun Komsuoglu$^{3}$,\\ Daniel I. Goldman$^{1}$\par}
{\centering\fontsize{9.3}{12.1}\selectfont
$^{1}$School of Physics, Georgia Institute of Technology, Atlanta, Georgia 30332, USA\\
$^{2}$Department of Mechanical Engineering, Northwestern University, Evanston, IL 60208, USA\\
$^{3}$Department of Electrical and Systems Engineering, University of Pennsylvania, Philadelphia, PA 19104, USA\par}
\vspace{4pt}
{\fontsize{9.5}{12.825}\selectfont
Published in Experimental Mechanics 50(9), 1383--1393 (2010).\\
DOI: \href{https://doi.org/10.1007/s11340-010-9347-1}{\underline{10.1007/s11340-010-9347-1}}\par}

\begin{abstract}
Achieving effective locomotion on diverse terrestrial substrates can require subtle changes of limb kinematics. Biologically inspired legged robots (physical models of organisms) have shown impressive mobility on hard ground but suffer performance loss on unconsolidated granular materials like sand. Because comprehensive limb-ground interaction models are lacking, optimal gaits on complex yielding terrain have been determined empirically. To develop predictive models for legged devices and to provide hypotheses for biological locomotors, we systematically study the performance of SandBot, a small legged robot, on granular media as a function of gait parameters. High performance occurs only in a small region of parameter space. A previously introduced kinematic model of the robot combined with a new anisotropic granular penetration force law predicts the speed. Performance on granular media is maximized when gait parameters minimize body acceleration and limb interference, and utilize solidification features of granular media.
\end{abstract}

\section{Introduction}\hypertarget{sec1}{}
To move effectively over a wide range of terrestrial terrain requires generation of propulsive forces through appropriate muscle function and limb kinematics~\underline{\cite{alexanderbook,dicAfar}}. Most biological locomotion studies have focused on steady rhythmic locomotion on hard, flat, non-slip ground. On these surfaces kinematic (gait) parameters like limb frequency, stride length, stance and swing durations, and duty factor can change as organisms walk, run, hop and gallop~\underline{\cite{alexanderbook}}. There have been fewer biological studies of gait parameter modulation on non-rigid and non-flat ground, although it is clear that gait parameters are modulated as the substrate changes during challenges like climbing~\underline{\cite{golAche06,jayAirs99}}, running on elastic/damped substrates~\underline{\cite{ferAlia99}}, transitioning from running to swimming~\underline{\cite{bieAgil99}}, and running on different preparations of granular media~\underline{\cite{golAkor06}}. Even subtle kinematic changes in gait can lead to major differences in limb function~\underline{\cite{autAhsi06}}. A major challenge is to develop models of limb interaction with complex substrates and to develop hypotheses for how organisms vary gait parameters in response to substrate changes.

The RHex class of model locomotors (robots) has proved useful to test hypotheses of limb use in biological organisms on hard ground~\underline{\cite{holAful06}} and recently on more complex ground with few footholds~\underline{\cite{spaAgol}} or the ability to flow~\underline{\cite{cheAumb09}}. These hexapedal devices model the dynamically stable locomotion of a cockroach and were the first legged machines to achieve autonomous locomotion at speeds exceeding one body length/s. In these devices, complexity in limb motion is pared down to a few biologically relevant parameters controlling intra-cycle ``stance'' and ``swing'' phases of 1-dof rotating limbs (referred to as``gait'' parameters hereafter; see detailed description in Methods and Results). When these gait parameters are appropriately adjusted, RHex shows performance comparable in speed and stability to organisms on a diversity of terrain~\underline{\cite{schAgar}}. However, because of the scarcity of existing models of limb interaction with complex substrates, adjustment of the gait parameters is typically done empirically~\underline{\cite{weiAlop04,koditschek04}}.

Sand, a granular medium~\underline{\cite{jaeAnag}}, is of particular interest for studies examining the effects of limb kinematics on locomotor performance on yielding terrain. In a previous study~\underline{\cite{cheAumb09}} we found that minor changes in the limb kinematics of a small RHex-class robot, SandBot (Fig.~\hyperlink{fig1}{\underline{1a}}), produced major changes in its locomotor mode and performance (speed) on a granular medium, poppy seeds (see Section \hyperlink{sec2.2}{\underline{2.2}}). This sensitivity occurs, in part, because forced granular media remain solid below the yield stress, but can flow like a fluid when the yield stress is exceeded~\underline{\cite{neddermanbook}}. We tested SandBot on granular media of different yielding properties (set by granular volume fraction; see Section \hyperlink{sec2.1}{\underline{2.1}}) at various limb frequencies but with the other gait parameters fixed. While there is no fundamental theory at the level of fluid mechanics that accounts for the physics of the solid-fluid transition of granular media or the dynamics of the fluidized regime, empirical models of granular penetration force have proved useful to predict SandBot's speed~\underline{\cite{cheAumb09}}. SandBot's propulsion is determined by factors that control this transition during limb-ground interaction (limb penetration depth, limb speed, body mass, grain friction, volume fraction, etc.). Using a simplified equation describing the granular penetration force, we developed a kinematic model to explain the locomotion of SandBot (see Section \hyperlink{sec2.2}{\underline{2.2}}).

In this study, we advance our understanding of the effects of limb kinematics on locomotor performance by testing SandBot with varying gait parameters on sand of fixed yield strength and at fixed limb frequency. We find that robot speed depends sensitively on limb kinematics; while the original model qualitatively captures this sensitivity, the penetration force used in the model and other assumptions need to be modified to explain some important features. Our study not only reveals the specific optimal kinematics for SandBot on granular media, but also advances our understanding of how in general to achieve effective legged locomotion on complex terrain.

\section{Background and Review of Previous Study}\hypertarget{sec2}{}

To understand the effect of limb kinematics addressed here, we first summarize the mechanism of SandBot locomotion on granular media (called rotary walking) discovered in our previous study~\underline{\cite{cheAumb09}}. In this section, we discuss the physics of granular media that controls the limb penetration depth (which governs locomotion performance) and then review our previous experiments and kinematic model.

\subsection{Physics of Limb-Granular Media Interaction}\hypertarget{sec2.1}{}

The physics that controls locomotor performance is the relative magnitude of the penetration resistance force (originating in the granular media) and the sum of the external forces (weight, inertial forces). When these balance, the granular media solidifies, allowing the robot to be supported at a fixed limb penetration depth.

The previous SandBot study~\underline{\cite{cheAumb09}} revealed that as the limb (or any simple intruder) vertically penetrates into the medium, the penetration force scales with $z$, the depth of the intruder below the surface~\underline{\cite{hilAyeu05}}, as $F_p(z) = k(\phi)z$, where $\phi$ is the volume fraction, the ratio of the solid volume of the granular media to the volume that it occupies (for natural dry sand, $0.55 < \phi < 0.64$). The constant, $k(\phi)$, characterizes the penetration resistance and increases with $\phi$. In this paper we keep $\phi$ fixed at approximately the critical packing state~\underline{\cite{umbAgol09,neddermanbook,schofieldbook}} (which is close to the as-poured volume fraction) where granular media neither globally dilate nor compact in response to shear (see Section \hyperlink{sec3}{\underline{3}}).

\manuscriptfigure{1}{width=\linewidth,height=4.6in,keepaspectratio}{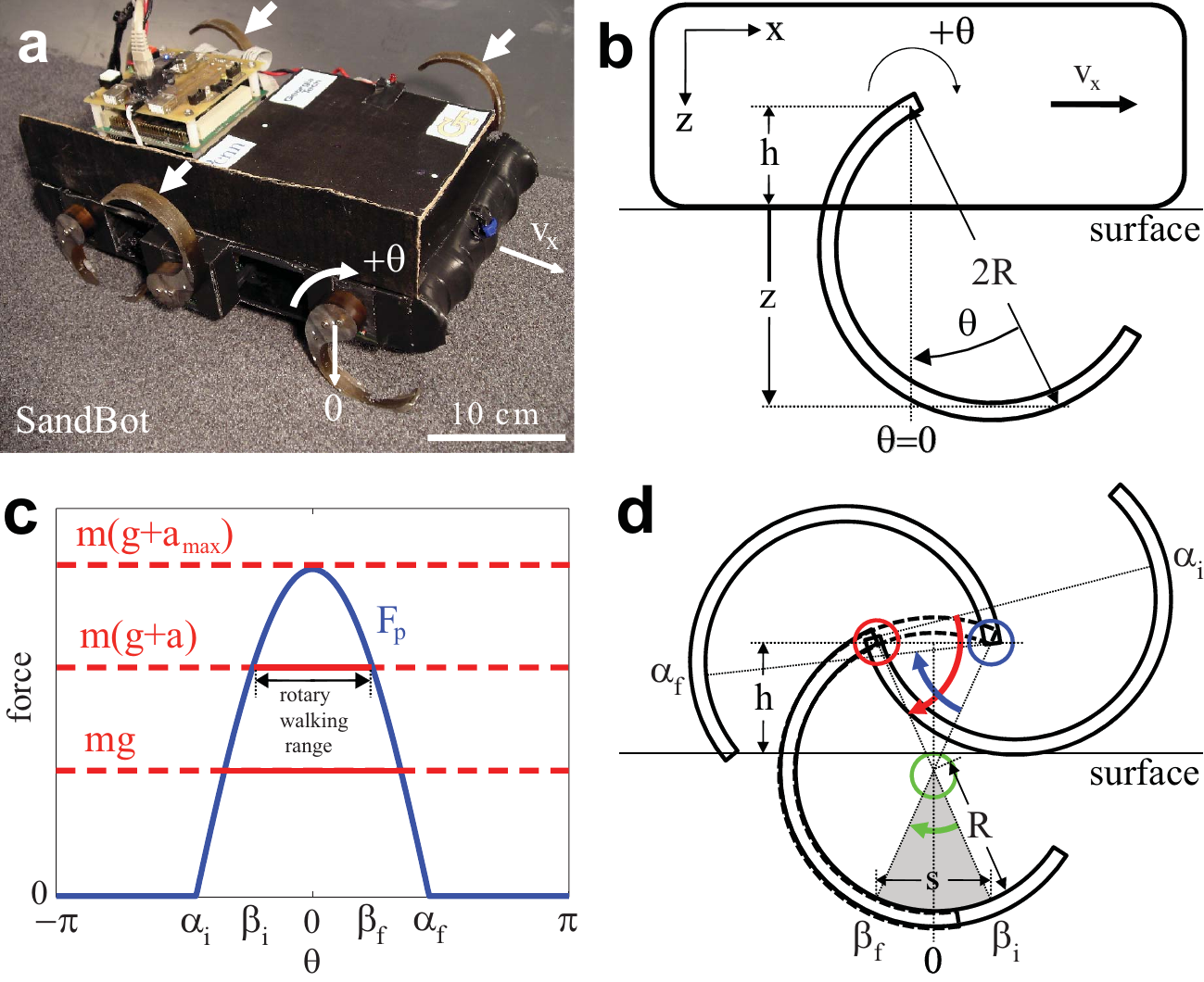}{
Mechanism of SandBot locomotion on granular media. (a) SandBot, a six-legged insect inspired robot, moves with an alternating tripod gait. The three arrows indicate the limbs of one tripod. (b) Schematic of single-leg representation of SandBot with mass $m = 1/3$ SandBot's total mass. With the body contacting the surface, the motor axle is height $h=2.5$~cm above the ground. The c-leg is approximately a circular arc (radius $R=3.55$~cm, arc span $225$ degree). Leg angle $\theta$ is measured clockwise about the axle and between the downward vertical and a diameter through the axle. Leg depth $z = 2R\cos\theta-h$. (c) Magnitude of penetration force $F_p$ (blue curve) relative to force required for upward motion, $m(g+a),$ (red lines) determines the locomotor mode. The force required for quasi-static movement ($mg$) is shown for reference. When $F_p$ and $m(g+a)$ intersect rotary walking occurs. When $F_p$ and $m(g+a)$ do not intersect (dashed red curve and above), the robot swims. (d) Schematic of rotary walking. The granular material flows in the intervals [$\alpha_i$, $\beta_i$] (red arrow) and [$\beta_f$, $\alpha_f$] (blue arrow) where $F_p < m(g+a)$ and the c-leg rotates about the axle (red and blue circles). The material is a solid in the interval [$\beta_i$, $\beta_f$] (gray sector; line with arrows in (c)) where $F_p$ exceeds $m(g+a)$ and the c-leg rotates about its center (green circle and arrow), lifting and propelling body forward by step length $s = R(\sin\beta_f-\sin\beta_i)$. The c-leg is above the ground in the interval [$\alpha_f$, $\alpha_i + 2\pi$]. Note that $[\alpha_i, \alpha_f]$ in (c) is symmetric to vertical ($\theta = 0$) as a result of assuming the force is isotropic (see Section \hyperlink{sec4.2}{\underline{4.2}}).
}

\subsection{Review of Previous Observations and Model}\hypertarget{sec2.2}{}

In the previous study of SandBot~\underline{\cite{cheAumb09}}, we fixed intra-cycle limb kinematics (by using gait parameters that produce consistent motion on granular media, see Section \hyperlink{sec3}{\underline{3}}) and measured SandBot's average speed $\overline{v}_x$ on poppy seeds as a function of volume fraction and the cycle-averaged limb frequency $\omega.$ We observed a sensitive dependence of $\overline{v}_x$ on both $\phi$ and $\omega$, and developed a kinematic model which explained this dependence and revealed two distinct locomotor modes determined by whether the granular media solidifies during limb-ground interaction.

Our kinematic model describes the limb-ground interaction of SandBot by considering the motion of just a single limb (Fig.~\hyperlink{fig1}{\underline{1b}}). SandBot has six approximately c-shaped limbs (c-legs) divided into two alternating tripods.  C-legs in the same tripod rotate in synchrony and each c-leg rotates about a horizontal axis normal to the robot body. We simplify the multi-leg ground interaction of each tripod to that of a single limb carrying $1/3$ the total body mass $3m$ ($2.3$ kg), as body weight is approximately uniformly distributed between each c-leg. We also considered a c-leg as a simple intruder ignoring its more complicated geometry, {\em i.e.} $F_p(z) = kz.$ The previous study~\underline{\cite{cheAumb09}} showed that the simple intruder approximation gave approximately the same results as a more realistic treatment in which penetration force was integrated over the submerged leading surface of a c-leg. In this study we use the simple intruder approximation.

We make the approximation that SandBot's body is in stationary contact with the surface at the onset forward motion in each cycle~\underline{\cite{note2}}, and define the c-leg's angular position, $\theta,$ as the clockwise angular displacement from the configuration where the center of curvature of the c-leg is directly beneath the axle, see Fig.~\hyperlink{fig1}{\underline{1b}}. During a full rotation, as $\theta$ changes from $-\pi$ to $\pi$, the c-leg initially contacts the ground at $\theta=\alpha_i$ and loses ground contact at $\theta=\alpha_i.$  Because leg depth can be approximated as $z = 2R\cos\theta-h$ when the body is in contact with the surface~\underline{\cite{note1}}, penetration force can be written as $F_p(\theta) = 2Rk\left[\cos\theta-\frac{h}{2R}\right]$ (blue curve in Fig.~\hyperlink{fig1}{\underline{1c}}), where $R = 3.55$~cm and $h = 2.5$~cm are the radius of the c-leg and the hip height ({\em i.e.} distance from c-leg axle to underside of body) respectively.

Of prime importance in determining SandBot's performance is the magnitude of the penetration force $F_p(\theta)$ relative to the sum of the forces required to the support the body weight and accelerate the body upward $m(g+a)$ (red curve in Fig.~\hyperlink{fig1}{\underline{1c}}), where $g$ is the acceleration due to gravity and $a$ the acceleration~\underline{\cite{cheAumb09}}.  The relevant acceleration is given by the jump in robot speed when the granular media solidifies, $R\omega,$ divided by the characteristic response time of the c-leg interacting with the granular media, $\Delta t(\phi),$ {\em i.e.} $a=R\omega/\Delta t.$  Two distinct locomotor modes are possible depending on whether or not $F_p(\theta=0) > m(g+a)$:

\begin{enumerate}
\item Rotary walking -- movement with solidification (see Fig.\hyperlink{fig1}{\underline{1b-d}}): As the c-leg rotates into the ground after initial leg-ground contact at $\theta=\alpha_i$, the penetration force increases with increasing depth. In the rotary walking regime the material beneath the c-leg solidifies and leg penetration stops at an angle $\theta=\beta_i$ when $F_p(\beta_i)=m(g+a),$ see Fig.~\hyperlink{fig1}{\underline{1c,d}}. Since the frictional force between the c-leg and granular material is insufficient for the leg to roll, the c-leg instead rotates about its center of curvature (green circle and arrow) lifting and advancing the robot in the process.  Rotary walking continues until $\theta=\beta_f,$ beyond which the c-leg again penetrates through the material since $F_p(\theta)<m(g+a)$ and the body is again in contact with the ground (blue circle and arrow). Rotary walking thus occurs over a finite range of leg angle $\beta_i < \theta < \beta_f$ or $[\beta_i, \beta_f]$ (horizontal arrow in Fig.~\hyperlink{fig1}{\underline{1c}} and gray sector in Fig.~\hyperlink{fig1}{\underline{1d}}) where $\beta_i$ and $\beta_f$ are determined by $F_p(\beta_{i,f}) = 2Rk(\cos\beta_{i,f}-\frac{h}{2R}) = m(g+a).$ For a given $[\beta_i, \beta_f]$, Fig.~\hyperlink{fig1}{\underline{1d}} shows that the robot advances a distance $s = R(\sin\beta_f-\sin\beta_i),$ where we call $s$ the step size.  During one complete gait cycle of period $T,$ each alternating tripod advances the robot by $s,$ giving an average robot speed of $\overline{v}_x = 2s/T = s\omega/\pi$.

\item Swimming -- movement without solidification: When $F_p(0) < m(g+a)$ (Fig.~\hyperlink{fig1}{\underline{1c}}, dashed red curve), the granular material beneath the penetrating c-leg never solidifies and rotary walking does not occur, {\em i.e.} $\beta_i = \beta_f = 0.$  Instead, the limb constantly slips through the surrounding fluidized granular material, similar to a swimmer's arm in water, and the robot advances slowly ($\overline{v}_x < 1$~cm/s).  In this regime forward motion occurs when the frictional and inertial (drag) forces generated by the c-legs exceeds the frictional force between the robot body and the surface.
\end{enumerate}

The two constants characterizing the interaction with the granular medium, $k$ and $\Delta t$, together with limb frequency $\omega$, determine the relative magnitudes of $F_p$ and $m(g+a)$ and consequently control which locomotor mode the robot operates in. Reducing $k$ (by decreasing $\phi$) and/or increasing $\omega$ reduces the rotary walking range; in other words, the less compact the granular material is and/or the faster the limbs rotate, the deeper the c-legs have to penetrate before the granular material solidifies and rotary walking begins, and the more susceptible the robot is to entering the slow swimming mode. This simple kinematic model captures the observed sensitive dependence of $\overline{v}_x$ on $\phi$ and $\omega$, with $k(\phi)$ and $\Delta t(\phi)$ as two fitting parameters.

In summary, our previous study of SandBot~\underline{\cite{cheAumb09}} showed that to locomote effectively on granular media, limbs kinematics that access the solid phase of granular media should be employed.

\section{Methods and Results}\hypertarget{sec3}{}

The limb kinematics of each tripod during one cycle are parameterized by three ``gait parameters", see Fig.~\hyperlink{fig2}{\underline{2}}(a,b). The kinematics of both tripods are periodic (with period $T$) and offset by half a period $T/2$ but are otherwise identical. For the conditions in this and previous experiments, a motor controller in the robot ensures that the target kinematics are achieved. Limb kinematics consist of a ``swing" phase (orange), which is typically faster, and a ``stance" phase (green), which is typically slower, with respective frequencies $\omega_f$ and $\omega_s$.

During hard ground locomotion in the RHex-class of Robots (and for animal locomotion in general), "swing" and "stance" phases typically correspond to off-ground and ground-contact phases, respectively. But because during locomotion on granular media this correspondence is not necessarily true, we simply call them fast and slow phases. In practice the fast and slow phases are implicitly defined by the triplet $\{\theta_s, \theta_0, d_c\}$ where $\theta_s$ is the angular extent of the slow phase, $\theta_0$ is the angular location of the center of the slow phase, and $d_c$ is the duty cycle of the slow phase (the fraction of the period in the slow phase). Specifying the cycle averaged limb frequency $\omega$ fully determines the motion of the limbs in the robot frame. By definition, $\omega_s = \frac{\theta_s}{Td_c}$, $\omega_f = \frac{2\pi-\theta_s}{T(1-d_c)}$, and $\omega = \frac{2\pi}{T}$. Typically, gait parameters are set so that $\omega_s < \omega < \omega_f$, but the reverse is possible when $\theta_s$ becomes large enough and/or $d_c$ small enough.

In the first tests of SandBot on granular media (Fig.~\hyperlink{fig2}{\underline{2c}}), we found that kinematics tuned for rapid stable bouncing motion on hard ground (HGK: $\{\theta_s, \theta_0, d_c\} = \{0.85, 0.13, 0.56\}$) produced little motion on granular media (red curve). Empirical adjustment to soft ground kinematics (SGK: $\{\theta_s, \theta_0, d_c\} = \{1.10, -0.50, 0.45\}$) restored effective (walking) locomotion on granular media (blue curve).

In the previous study~\underline{\cite{cheAumb09}}, we used SGK to test $\overline{v}_x(\phi, \omega)$. Now armed with the understanding of how SandBot moves on granular media gained from this work, we set out to determine the effects of limb kinematics in detail. We set $\phi = 0.605$ and $\omega = 8$ rad/s, and measure SandBot's average speed on granular media as we systematically vary gait parameters, i.e. $\overline{v}_x$ = $\overline{v}_x(\theta_s, \theta_0, d_c)$. We pick $\omega = 8$ rad/s because at this intermediate frequency SandBot displays both rotary walking and swimming as the clock signal is varied. We pick $\phi = 0.605$ to remove the effect of local volume fraction change which causes a premature transition from rotary walking to swimming~\underline{\cite{note3}} and adds to the complexity of the problem.

We first test the effect of the extent and location of the slow phase for fixed $d_c = 0.5$, measuring speed $\overline{v}_x$ = $\overline{v}_x(\theta_s, \theta_0)$. We vary the parameters between $0 \leq \theta_s \leq 2$ and $-2 \leq \theta_0 \leq 2$, which are the limits set by the robot's controller. We choose $d_c$ = 0.5 because it is close to the $d_c$ values of both HGK and SGK. This gave us an easy way to project HGK ($d_c = 0.56$) and SGK ($d_c = 0.45$) onto the $\overline{v}_x$ = $\overline{v}_x(\theta_s, \theta_0)$ plot ($d_c = 0.5$), assuming that a small change of $d_c$ near $d_c=0.5$ does not affect speed significantly (see Fig.~\hyperlink{fig6}{\underline{6a}} and Section \hyperlink{sec4.3}{\underline{4.3}} which support this assumption)).

\manuscriptfigure{2}{width=\linewidth,height=4.6in,keepaspectratio}{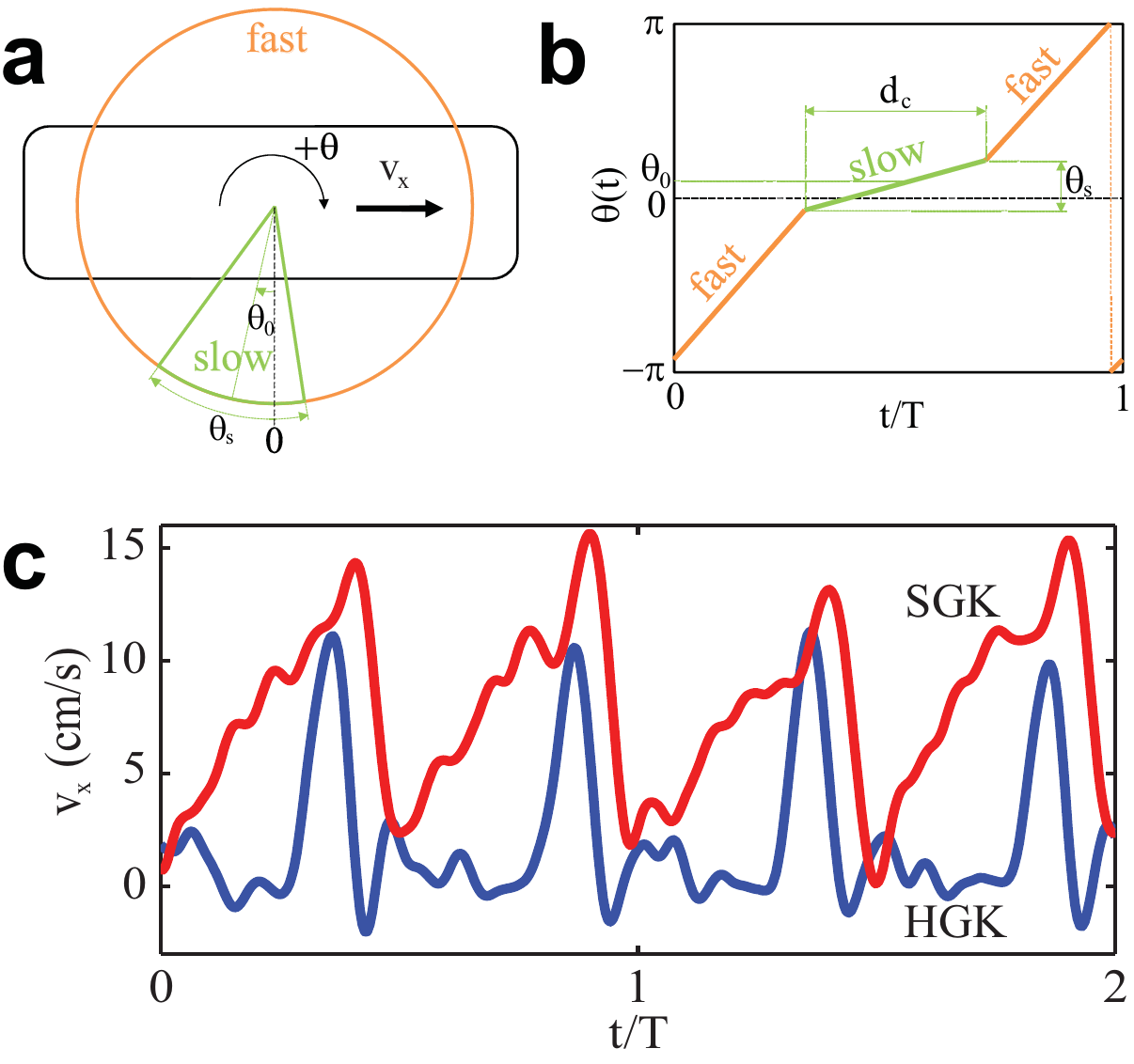}{
SandBot's intra-cycle limb kinematics and affects its speed on granular media. (a) Each leg rotation is composed of a fast phase (orange) and a slow phase (green). $\theta_s$ and $\theta_0$ define the angular extent and center of the slow phase respectively. (b) Leg angle $\theta$ as a function of time during one cycle (normalized to $T$). $\theta(t)$ of the other tripod is shifted by $T/2$ but otherwise identical. $d_c$ is the duty cycle of the slow phase, {\em i.e.} fraction of the period spent in the slow phase.  (c) Instantaneous speed of SandBot on granular media with hard ground clock signal (HGK: $\{\theta_s, \theta_0, d_c\} = \{0.85, 0.13, 0.56\};$ red) and soft ground clock signal (SGK: $\{\theta_s, \theta_0, d_c\} = \{1.10, -0.50, 0.45\};$ blue). With HGK SandBot moves slowly ($\overline{v}_x \approx 2$~cm/s) on granular media, but with SGK (red), it advances rapidly ($\overline{v}_x \approx 8$~cm/s).
}

Measurements of $\overline{v}_x$ = $\overline{v}_x(\theta_s, \theta_0)$ (Fig.~\hyperlink{fig3}{\underline{3a}}) show a single sharp peak in speed near \{$\theta_s, \theta_0$\} = \{1.5, $-$0.5\}. High speeds only occur within a small island of $-1 < \theta_0 < 0$ and $\theta_s>0.5$ surrounding the peak; lower speeds fill the remainder of the space. The drop in speed is rapid as $\theta_0$ is varied away from the peak, and is less so when $\theta_s$ is varied away from the peak; this is also evident in cross sections through the peak (blue circles in Figs.~\hyperlink{fig3}{\underline{3a}} and Fig.~\hyperlink{fig6}{\underline{6a}}, respectively). Ignoring the effect of $d_c$, the SGK parameters (blue dot) lie close to the peak while the HGK parameters (red dot) are in the low speed region.  The optimal gait parameters which we found for SandBot locomotion on poppy seeds at $\phi = 0.605$ and $\omega = 8$~rad/s are: $\{\theta_s, \theta_0, d_c\} =\{1.5, -0.5, 0.55\}$. These gait parameters generate about $20\%$ higher speed than the previously used SGK parameters.

\manuscriptfigure{3}{width=\linewidth,height=4.6in,keepaspectratio}{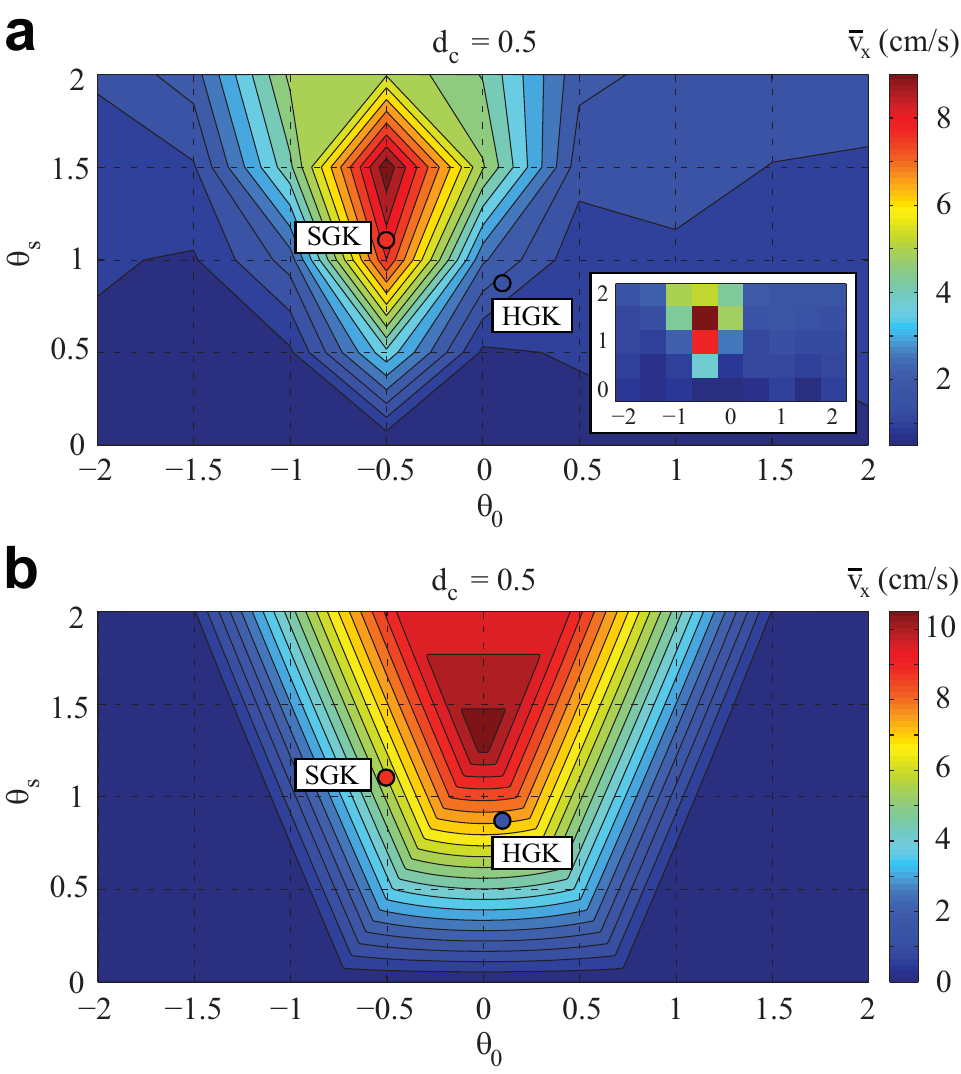}{
Average speed $\overline{v}_x$ of SandBot on granular media ($\phi= 0.605$) as a function of ($\theta_s$, $\theta_0$) for $d_c=0.5$ and $\omega= 8$~rad/s. (a) The experimental data has a localized region of high speeds with peak $\overline{v}_x \approx 9$~cm/s near $\{\theta_s, \theta_0$\} = \{1.5, $-0.5\}.$. Circles show that $\overline{v}_x$ for SGK (red) and HGK (blue) matches data for $d_c=0.5$ despite the formers slightly different $d_c$ values, see Section \hyperlink{sec4.3}{\underline{4.3}}. Inset: original data from which main figure is interpolated. (b) Predicted $\overline{v}_x(\theta_s, \theta_0)$ from the kinematic model with $F_p = kz$ captures the single peak but predicts a lower speed for SGK than for HGK contrary to observation and fails to account for the observed peak in speed at $\theta_0=-0.5$.
}

Variation of the duty cycle at fixed $\{\theta_s, \theta_0\} = \{1.5, -0.5\}$ also has a substantial influence on speed. Data (blue circles in Fig.~\hyperlink{fig6}{\underline{6b}}) show a well defined peak at $d_c \approx 0.5$. Speed drops off relatively slowly for $d_c >$ 0.55, and more quickly to small (swimming) speeds for $d_c < 0.5$.

\section{Discussion}\hypertarget{sec4}{}

\subsection{Application of Model to Slow Phase Extent and Location Variation}\hypertarget{sec4.1}{}

To apply our kinematic model to SandBot locomotion with varied limb kinematics, we must consider the effects of variable limb kinematics during limb-ground interaction. Depending on the gait parameters during ground contact, the limbs could be rotating in the fast phase, in the slow phase, or in a combination of both. In our previous study, the kinematic model ignored limb frequency variability during ground contact and only considered the robot limb rotating at the constant cycle averaged limb frequency $\omega$.

However, as limb kinematics change, the variability of limb frequency in ground contact needs to be taken into account. For our test of $\overline{v}_x$ = $\overline{v}_x(\theta_s, \theta_0)$ within $0 \leq \theta_s \leq 2$ and $-2 \leq \theta_0 \leq 2$ and at $d_c = 0.5$, $\omega_f >> \omega_s$ so that only the slow phase can possibly achieve rotary walking, as fast limb rotation results in swimming. In this case $\omega_s$ (instead of $\omega$) controls acceleration $a$ and thus determines the rotary walking range~\underline{\cite{note4}}.

For fixed $d_c$, varying [$\theta_s, \theta_0$] changes the extent and location of the slow phase; the angular extremes of the slow phase are $\theta_{i,f} = \theta_0 \pm \frac{\theta_s}{2},$ see Fig.~\hyperlink{fig5}{\underline{5a}}.  Varying $\theta_s$ also changes $\omega_s$ which controls the rotary walking range. Therefore varying [$\theta_s, \theta_0$] affects where the slow phase overlaps with the rotary walking range. The step length $s$ is given by $s = R(\sin \psi_f - \sin \psi_i)$ where $\psi_i = \max(\beta_i,\theta_i)$ and $\psi_f = \min(\beta_f,\theta_f)$ if there is overlap or $s = 0$ if there is no overlap. The larger the overlap, the further the robot moves forward in a cycle.

As shown in Section \hyperlink{sec2.2}{\underline{2.2}}, the rotary walking range [$\beta_i, \beta_f$] is given by solving the equation $F_p(\theta) = 2Rk(\cos\beta_{i,f}-\frac{h}{2R}) = m(g+a)$, with $a$ given by $a = \frac{m\omega_s}{\Delta t}$. We can evaluate how [$\theta_i,\theta_f$] overlaps with [$\beta_i, \beta_f$] to determine $s$, and calculate the robot speed using $\overline{v}_x = \frac{2s}{T} = \frac{s\omega}{\pi}$.  For fixed $\omega$, speed $\overline{v}_x$ scales with step length $s$.

Figure~\hyperlink{fig3}{\underline{3b}} shows the model prediction of $\overline{v}_x$ using fitting parameters $k=210$~N/m, $\Delta t = 0.37$~s. Comparing prediction with observation (Fig.~\hyperlink{fig3}{\underline{3a}}), the model captures the peak and predicts similar magnitudes of speeds. However the predicted peak is symmetric about $\theta_0=0$ while the observed peak is symmetric about $\theta_0 = -0.5.$

\subsection{Anisotropic Penetration Force Law}\hypertarget{sec4.2}{}

If the penetration force of the granular material increased like $kz$ as assumed in the model, we would expect $\theta_0 = 0$ as this value would give the largest overlap between the slow phase and the rotary walking range as determined by the material strength (see scheme in Fig.~\hyperlink{fig5}{\underline{5a}}). To investigate why the robot performs best with $\theta_0= -0.5$ we attached a c-leg to a force/torque sensor and measured the grain resistance as the c-leg was rotated through the granular media at $\omega=0.35$~s$^{-1}$ (the horizontal rotation axis of the c-leg was positioned the same distance $h=2.5$~cm above the grain surface as when it is mounted on the robot). Figure~\hyperlink{fig4}{\underline{4b}} shows a clear asymmetry in the penetration (vertical) force with the measured peak force occurring near $\beta_0 = -0.75$; penetration force during rotation peaks before the intruder reaches the maximum depth. We confirmed that the measured anisotropy in the penetration force is intrinsic to our granular medium and is not an artifact of the particular shape of the c-leg by additionally rotating a rectangular bar and a sphere into granular media at the same hip height: both objects exhibited a peak force at $\beta_0 = -0.75$.

We speculate that the asymmetry in penetration force during rotation into granular media is a result of the changing limb orientation during rotational intrusion. For vertical penetration (which we considered in the model in the previous study), the intruder is constantly pushing down on the granular material. The grain contact network generated in granular material in response to intrusion~\underline{\cite{GenAHow01}} forms a downward pointing cone which generates a force symmetric to the vertical ($\theta = 0$). In rotational intrusion, however, the direction of intrusion is constantly changing; the direction of the force cone should change as well and correlate with the instantaneous direction of intrusion.

\manuscriptfigure{4}{width=\linewidth,height=4.6in,keepaspectratio}{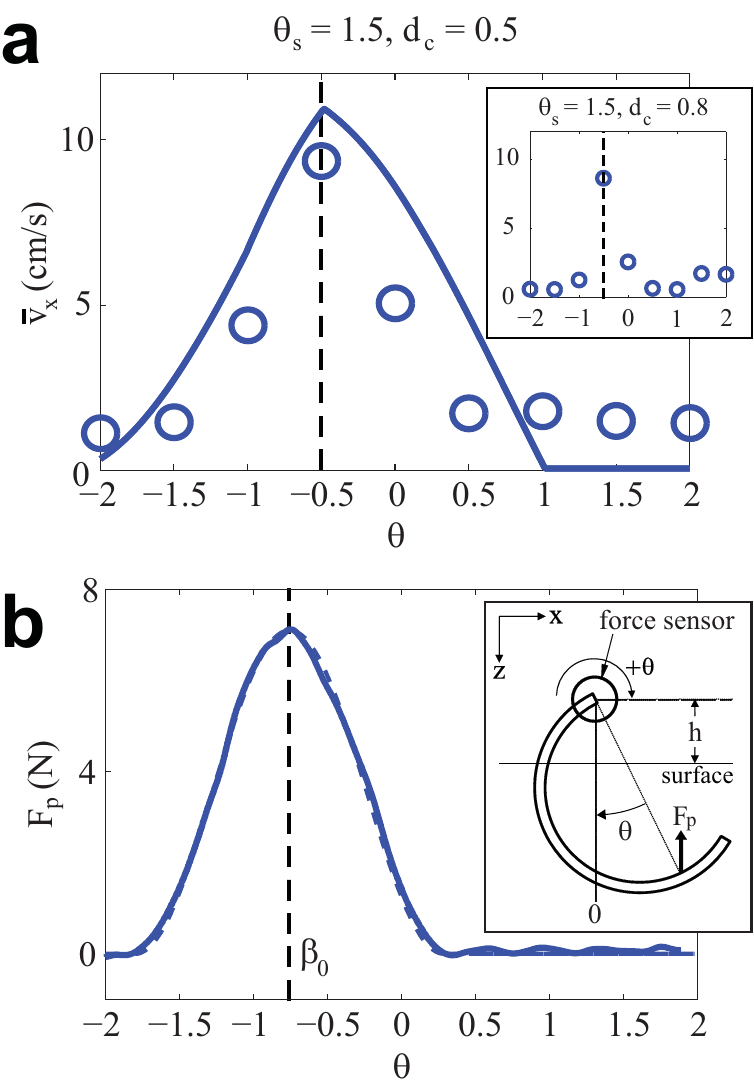}{
Asymmetry of $\overline{v}_x$ with respect to $\theta_0 = 0$ is due to anisotropic penetration force during limb rotation into granular media. (a) For all $\theta_s$ at $d_c=0.5,$ $\overline{v}_x(\theta, \theta_0)$ (Fig.~\hyperlink{fig3}{\underline{3a}}) is maximal (dashed vertical black line) at $\theta_0 = -0.5$ ($\theta_s=1.5$ shown). Inset: peak location $\theta_0 = -0.5$ does not change for $d_c = 0.8$. (b) Vertical penetration force $F_p$ (solid blue curve) during c-leg rotation into poppy seeds reaches maximum at $\beta_0 = -0.75$ (dashed black line) and is asymmetric to $\theta_0 = 0$. Inset: force measurement schematic. Solid blue curve in (a) is prediction from the model with anisotropic penetration force.
}

We hypothesize that the force during rotational intrusion is maximal at $\beta_0=-0.75$ because for larger angles part of the cone reaches the surface and/or terminates on the horizontal walls of the container and can no longer support the entire grain contact network, thus reducing the maximal yield force. We also note that the angle at which maximal force is developed is close to the angle of repose $0.52$ that we measure for the poppy seeds. This angle is the same as the internal slip angle in cohesionless granular material~\underline{\cite{neddermanbook}} which plays an important role in the formation of the grain contact network, supporting the plausibility of our speculation.

To account for the measured angular offset in peak force from vertical (Figs.~\hyperlink{fig2}{\underline{2c}} and 3a), we modify the original penetration force law in our model to $$F_p(\theta) = 2Rk'\left\{\cos\left[b\left(\theta-\beta_0\right)\right]+1\right\}$$ for $F>0$, where $\beta_0 = -0.75$, and $k'$ and $b$ are new fit parameters. Following the same procedure described in Section \hyperlink{sec4.1}{\underline{4.1}}, we find the robot speed by calculating [$\beta_i, \beta_f$], the overlap between [$\theta_i,\theta_f$] and [$\beta_i, \beta_f$], and the step size.

Figure~\hyperlink{fig5}{\underline{5c}} shows $\overline{v}_x$ predicted by a fit to the model using the anisotropic penetration force law (fitting parameters $k' = 65$~N/m, $\Delta t = 0.4$~s, and $b = 0.8$). Besides capturing the peak behavior of measured speed, the model also captures the shift in peak location to $\theta_0 = -0.5$ ({\em i.e.} asymmetry to $\theta_0 = 0$). For fixed $\theta_s$, speed is maximal when the center of the slow phase corresponds with the center of the rotary walking range (Fig.~\hyperlink{fig5}{\underline{5a}}). If $\theta_0$ is different from $\beta_0 = -0.75$, the overlap of the slow phase and the rotary walking range decreases, which reduces step length and thus speed. In accord with observation, SGK (red dot) lies near the peak while HGK (blue dot) lies in a region of low speeds. Figure~\hyperlink{fig5}{\underline{5a,b}} shows that SGK has higher speed than HGK because the overlap between the slow phase and the rotary walking range (gray sector in Fig.~\hyperlink{fig5}{\underline{5a}}) is significantly larger.

\manuscriptfigure{5}{width=\linewidth,height=4.6in,keepaspectratio}{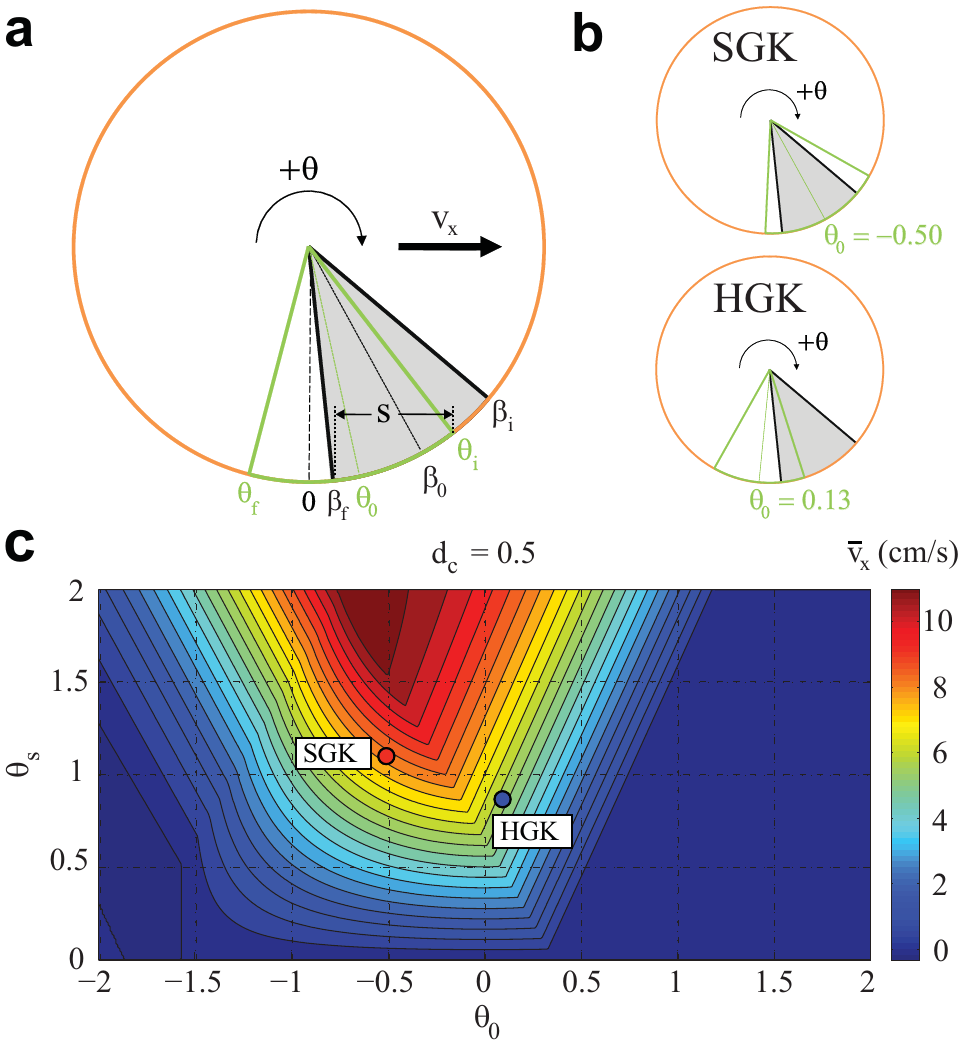}{
Overlap of the slow phase and the rotary walking interval predicted by the anisotropic penetration force law better predicts $\overline{v}_x(\theta_s, \theta_0)$. (a) Overlap of the slow phase $[\theta_i, \theta_f]$ and the rotary walking range $[\beta_i, \beta_f]$ determines step length $s$ and thus speed $\overline{v}_x$. For the configuration shown, $s = R(\sin\beta_f - \sin\theta_i)$. $\theta_0$ and $\beta_0$ are centers of the slow phase and the rotary walking range, respectively. (b) In SGK the slow phase (centered at $\theta_0 = -0.50$) overlaps nearly completely with the rotary walking range (centered at $\beta_0 = -0.75$), explaining its high speed as compared to HGK (centered at $\theta_0 = 0.13$) which has little overlap. (c) Prediction of $\overline{v}_x(\theta_s, \theta_0)$ from the the model with the anisotropic penetration force for $d_c = 0.5$ captures the asymmetry of $\overline{v}_x$ with respect to $\theta_0 = 0$ and predicts higher speed for SGK than for HGK. Fitting parameters: $k' = 65$~N/m,  $\Delta t = 0.4$~s and $b= 0.8$.
}

At fixed $\theta_0 = \beta_0$, increasing the extent of the slow phase (increasing $\theta_s$) from zero initially increases speed as the extent of the slow phase increases within the rotary walking envelope (see Figs.~\hyperlink{fig2}{\underline{2a}} and 6a). However, $\omega_s$ increases with $\theta_s$ which increases the acceleration and reduces the rotary walking range.  For sufficient extent (near $\theta_s = 1.5$ for the data shown in Figs.~\hyperlink{fig2}{\underline{2a}} and 6a) the slow phase contains the rotary walking range and step length is determined by the latter.  Further increase in $\theta_s$ reduce the rotary walking range. Rotary walking is not possible for $\theta_s \geq \frac{2\pi d_c \Delta t}{\omega}\left(\frac{4k'}{m}-\frac{g}{R}\right)$ as the material is never strong enough to both support and accelerate the robot. In Fig.~\hyperlink{fig6}{\underline{6a}} the experimental speed is noticeably lower than the model prediction at the largest $\theta_s = 2$. As we discuss below in regards to variation in $d_c$, this reduction is a apparently the result of tripod overlap (both tripods simultaneously in ground contact) which occurs for a greater portion of the slow phase for larger $\theta_s.$

\subsection{Effect of Duty Cycle}\hypertarget{sec4.3}{}

While the model prediction of $\overline{v}_x(d_c)$ (blue curve in Fig.~\hyperlink{fig6}{\underline{6b}}) matches the magnitudes of speeds at intermediate $d_c \approx 0.5$, it does not quantitatively match the shape of measured speed vs $d_c$. Below $d_c \approx 0.5$, the model predicts that $\overline{v}_x(d_c)$ increases monotonically with increasing $d_c$. This trend is in accord with the experimental observations; however the model prediction is consistently higher than measured speed. Above $d_c \approx 0.5$, the model predicts that $\overline{v}_x(d_c)$ is independent of $d_c$, but the measured speed decreases with increasing $d_c$ and is lower than the model prediction. We now discuss possible reasons for the discrepancies at low $d_c < 0.1$ (labeled region I in Fig.~\hyperlink{fig6}{\underline{6b}}), intermediate $0.1 < d_c < 0.5$ (region II) and high $d_c > 0.5$ (region III).

In region I, $d_c$ is small so that $\omega_s$ and thus $a$ become large enough to ensure that the robot is in the swimming mode (i.e. movement without solidification). Here the model assumes rotary walking and predicts zero speed. In experiment, the robot can still advance slowly at each step due to thrust forces from continuously slipping limbs generated by frictional drag~\underline{\cite{albert99}} and/or inertial movement of material. This thrust competes with friction from belly drag, and as in~\underline{\cite{cheAumb09}}, we find that these result in low average speed of $\overline{v}_x \approx 1$ cm/sec.

The model's overestimate of speed at high $d_c$ (region III) is a result of tripod overlap and can be readily understood.  When there is no tripod overlap (only one tripod with ground contact at any given time) each tripod advances the robot a distance  $s$ for a total displacement of $2s$ per period. This is the case for $d_c \leq 0.5$. However in the limit of $d_c = 1$ both tripods are simultaneously in the slow phase as the duration of the fast phase is zero. The simultaneous slow phases generate a total displacement of just $s$~\underline{\cite{note5}} instead of $2s$ without tripod overlap. As a result, the predicted speed at $d_c = 1$ must be halved (i.e. $\overline{v}_x = \frac{s}{T} = \frac{s\omega}{2\pi}$). Lacking a way to quantify the tripod overlap effect, we assume that the reduction of step length from $2s$ to $s$ is linear with $d_c$ for  $d_c > 0.5$; the data is in good agreement with this prediction (dashed blue curve in Fig.~\hyperlink{fig6}{\underline{6b}}). This reduction in speed is a purely kinematic effect that is inherent to the rotary walking gait at high $d_c$.

\manuscriptfigure{6}{width=\linewidth,height=4.6in,keepaspectratio}{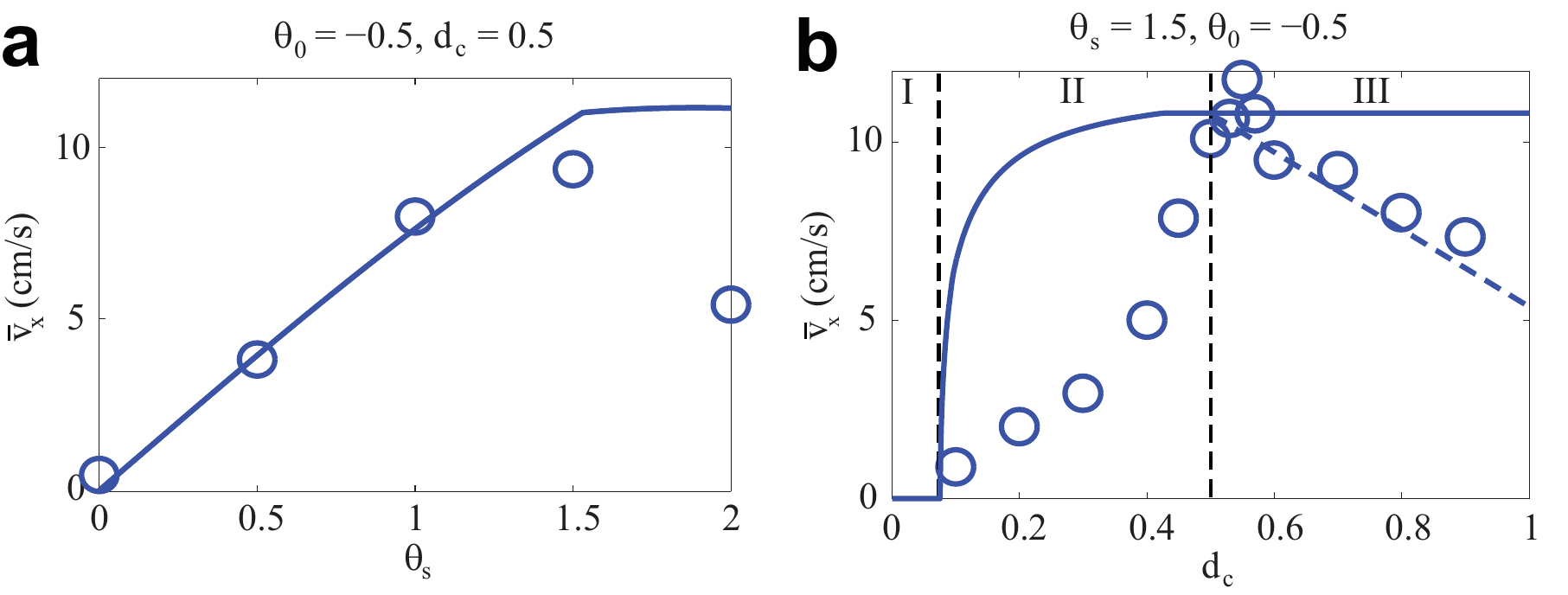}{
Comparison of data and anisotropic force model predictions for $\overline{v}_x(\theta_s)$ and $\overline{v}_x(d_c).$ (a) Measured $\overline{v}_x(\theta_s, \theta_0=-0.5)$ (blue circles) through the speed maximum (Fig.~\hyperlink{fig6}{\underline{6a}}) deviates from the anisotropic force model prediction (blue curve) at large $\theta_s$ due to limb overlap. (b) Measured $\overline{v}_x(d_c)$ (blue circles) for $\{\theta_s=1.5, \theta_0=-0.5\}$ is maximum at $d_{c} \approx 0.5$. The model (solid blue curve) accurately predicts the speed for $d_{c} \approx 0.5$ but is inaccurate elsewhere due to contributions of swimming neglected by the model (region I), a decrease in rotary walking range from cratering induced depth reduction and unequal penetration forces developed by c-legs on opposite sides of the body (II), and tripod overlap at high $d_c$ (III).  Model prediction with tripod overlap included (dashed blue line) better matches the data in region III.
}

Two plausible mechanisms explain the model's overestimate of speed at intermediate $d_c$ (region II): hole digging and uneven weight distribution. Lateral observations of the robot kinematics at low $d_c$ show that the rapid motion of the c-leg during the slow phase throws significant numbers of particles out of the limb's path which creates a depression. For a deep enough hole, rotary walking is impossible due to the reduced penetration depth of the leg below the now lower surface of the depression. The second mechanism concerns the model's assumption of uniform weight distribution between the three legs of the tripod. In the $d_c$ range where the robot advances slower than predicted, observations show that the robot rotates in the horizontal plane. Rotation occurs in this transition region between pure swimming and pure rotary walking because the side of the robot with two c-legs in ground contact undergoes rotary walking while the opposite side is in the swimming mode. Due to the increased gravitational and inertial forces on the single c-leg, the penetration forces are never sufficient to achieve ground solidification under the leg even at the maximum penetration depth.

\section{Conclusions}\hypertarget{sec5}{}

We have built upon our previous experiments and models of a legged robot, SandBot, to explore how changes in limb kinematics affect locomotion on granular media. We found that even when moving on controlled granular media of fixed volume fraction at fixed cycle-averaged limb frequency, speed remains sensitive to variations in gait parameters that control angular extent, angular location, and temporal duty factor of the slow phase of the limb cycle. We showed that the assumptions in a previously introduced model (which accurately predicted speed as a function of limb frequency and volume fraction) had to be modified to incorporate an anisotropic penetration force during rotational intrusion into granular media as well as changes in acceleration of the leg as gait parameters were varied. With these modifications the model was able to capture speed as a function of angular extent and angular location. The model also indicates that as duty cycle is changed, effects due to simultaneous limb pairs (tripods) in ground contact, rapid limb impact into sand, and unequal weight distribution on limbs become important.

Our experiments and modified model explain why gait parameters that allow the robot to rapidly bounce over hard ground lead to loss of performance on granular media. They demonstrate how the angular extent and location of the slow phase must be adjusted to optimize interaction with granular media by minimizing inertial force and limb interference, and maximizing the use of solid properties of granular media. Further studies of SandBot guided by our kinematic model should reveal how physical parameters of both robot (mass distribution, limb compliance, limb shape, belly shape) and the environment (grain friction, density, incline angle, gravity) control the solid-fluid transition and thus affect the limb-ground interaction and performance. However, advances are required in theory and experimental characterization of complex media. Otherwise we must continue to rely on empirical force laws specific to particular geometries, kinematics and granular media.

The existence of a speed optimum in gait parameter space implies that control of limb kinematics is critical to move effectively on granular media, whether actively through sensory feedback, and/or passively through mechanical feedback. Future work should compare these results to investigations of gait optimization on hard ground~\underline{\cite{weiAlop04}}. The differences in limb kinematics on sand compared to hard ground are intriguing because on hard ground performance is optimized by making the robot bounce. However, this carries with it the risk of yaw, pitch and roll instability due to mismanaged kinetic energy. On granular media such instabilities appear rare; instead most gait parameters (see Fig.~\hyperlink{fig3}{\underline{3a}}) result in little or no forward movement due to mismanaged fluidization of the ground. Thus, our results could have a practical benefit as they suggest strategies for improving the performance of current machines~\underline{\cite{hooAste,plaAbue06,marsrover}} on variable terrain via new limb and foot designs and control strategies.

Finally, an enormous number of organisms contend with sand~\underline{\cite{brownbook}}, moving on the surface (or even swimming within it~\underline{\cite{malAdin09}}).  While the observed phenomena and proposed locomotion modes (e.g. rotary walking) appear specific to SandBot and its c-shaped limbs, the underlying principles could apply to locomotion of organisms on yielding substrates. For example, our recent work on terrestrial hatchling sea turtle locomotion demonstrates that their effective movement on sand proceeds through solidification of the granular medium~\underline{\cite{mazAgra10}}. Integrated studies of biological organisms and physical models can provide hypotheses~\underline{\cite{fulAkod}} for passive and active neuro-mechanical~\underline{\cite{nisAbie07}} control strategies as well as better understanding of energetics~\underline{\cite{lejeune98}} for movement on complex terrain.

\section*{Acknowledgements}

We thank Daniel Koditschek, Ryan Maladen, Yang Ding, Nick Gravish, and Predrag Cvitanovi\'c for helpful discussion. This work was supported by the Burroughs Wellcome Fund (D.I.G., C.L., and P.B.U.), the Army Research Laboratory (ARL) Micro Autonomous Systems and Technology (MAST) Collaborative Technology Alliance (CTA) under cooperative agreement number W911NF-08-2-0004 (D.I.G. and P.B.U.), and the National Science Foundation (H.K.).


\begin{thebibliography}{10}
\bibitem{note2}
When tripods move in phase ({\em e.g.} $d_c=1$) this approximation is exact as
  the body rests on the surface during the swing phase.
\bibitem{note1}
In our previous study~\underline{\cite{cheAumb09}} we measured leg depth to the bottom of
  the c-leg; here we measure it to the point on the c-leg furthest from the
  motor axle to simplify the expression for penetration force vs.\ angle. As
  $mg$ is comparable to $kz$ for this study, the smallest penetration depth
  where rotary walking begins is close to the maximum possible value of $2R-h$
  so that the simplified expression for the leg depth is nearly identical to
  the exact value.
\bibitem{note3}
In rotary walking when $s < R$, a limb encounters material disturbed by its
  previous step. If the initial volume fraction exceeds the critical value
  $\phi\approx 0.605$, disturbed granular material dilates to a lower volume
  fraction after each step; if initially $\phi<0.605$, disturbed granular
  material compacts to a higher volume fraction after each
  step\underline{\cite{umbAgol09}}. The dilation of disturbed granular material above
  $\phi\approx 0.605$ results in premature transition from rotary walking to
  swimming if $s < R$ as the disturbed granular material is weaker which
  increases penetration and reduces step length. Here we choose $\phi = 0.605$
  which ensures that the $\phi$ encountered by the leg is unchanged even for $s
  < R$.
\bibitem{note4}
The previous study did not err in considering $\overline{v}_x$ as a function of
  $\omega$ alone since for fixed clock parameters $\omega_s$ scales with
  $\omega$. The only difference would be $\Delta t(\phi) \rightarrow
  \omega_s/\omega \Delta t.$.
\bibitem{note5}
For $d_c = 1$, tripods are in phase so that each c-leg needs to provide just
  1/6 the required total force compared to 1/3 when the tripods act
  independently which could affect the step size. However, we found that for
  $\{\theta_s, \theta_0, d_c\} = \{1.5, -0.5, 1\}$, decreasing $m$ from 1/3 to
  1/6 the body mass left the expression for $s$ unchanged.
\end{thebibliography}
\end{document}